\documentclass{webofc}
\usepackage[varg]{txfonts} 
\usepackage{color}
\usepackage{array}
\newcommand{\mbx}{\mathbf{x}}
\newcommand{\mbk}{\mathbf{k}}
\newcommand{\mbp}{\mathbf{p}}

\begin{document}
\title{Event-by-event Kinetic Description of Pre-equilibrium Charge Evolution in QCD Plasma}

%
%

\author{\firstname{Travis} \lastname{Dore}\inst{1}\fnsep\thanks{\email{tdore@physik,uni-bielefeld.de}} \and
        \firstname{Xiaojian} \lastname{Du}\inst{1,2}\fnsep \and
        \firstname{Sören} \lastname{Schlichting}\inst{1}\fnsep
}

\institute{Fakult\"at f\"ur Physik, Universit\"at Bielefeld, D-33615 Bielefeld, Germany 
\and
    Instituto Galego de Física de Altas Enerxías (IGFAE), Universidade de Santiago de Compostela, E-15782 Galicia, Spain          }

\abstract{%
We use QCD effective kinetic theory to calculate far-from-equilibrium dynamics on an event-by-event basis within the K\o{}MP\o{}ST framework. We present non-equilibrium charge response functions and the dynamical evolution of the conserved charge current pertinent to the early-time dynamics of heavy-ion collisions at the highest energies. The K\o{}MP\o{}ST framework with conserved baryon, strangeness, and electric charges can then be readily implemented into a multistage model allowing for the initialization of a non-equilibrium charge current in hydrodynamic simulations.
}
\maketitle
\section{Introduction}
\label{intro}
It is well known that at the earliest times in heavy-ion collisions, the system begins in a state that is very far from equilibrium, while at later times, the system lends itself well to a hydrodynamic description~\cite{Luzum:2008cw,Giacalone:2017dud,ALICE:2015juo}. The K\o{}MP\o{}ST framework is a tool which solves non-equilibrium linear response for heavy-ion systems and is able to take the system from the far-from-equilibrium initial state to the local equilibrium hydrodynamic state in a physical manner~\cite{Kurkela:2018wud,Kurkela:2018vqr}. In the precision era of heavy-ion physics, it is important to continue updating our models to capture more of the relevant physics. One such aspect that has been left out of state-of-the-art models is the evolution of conserved charges in the system at the highest LHC energies. Experimental results have shown that, even at the highest energies of RHIC and LHC, there is some finite net-baryon number deposited at mid-rapidity~\cite{ALICE:2019nbs}. Recent theoretical work has constructed a framework which is able to produce this effect~\cite{Garcia-Montero:2023gex}. In this proceedings, we show some results after having upgraded the K\o{}MP\o{}ST framework to include the evolution of conserved charges. Since the charge sector is new, we focus entirely on these results.

\section{QCD Kinetic Theory and Response Functions}
\label{KT}
In this work, we employ the QCD effective kinetic theory ( EKT)~\cite{Du:2020dvp,Du:2020zqg} which includes inelastic and elastic collision kernels computed in the AMY formalism \cite{Arnold:2002zm}. Within the K\o{}MP\o{}ST framework~\cite{Kurkela:2018wud,Kurkela:2018vqr}, the longitudinally boost-invariant expansion of the bulk of the system is treated as the dominant effect. We refer to this portion of the system as the background which is isotropic in the transverse plane. On the other hand, we also include the effect of spatial fluctuations by introducing a perturbation on top of this background which is not isotropic in the transverse plane. All together, we solve the following coupled set of kinetic theory equations
\begin{eqnarray}
\left(\partial_{\tau}
-\frac{p_{\|}}{\tau}\partial_{p_{\|}}\right) \bar{f}_a
=-C_a[\bar{f}]\;,~~~~~~
\left(\partial_{\tau} + i\frac{\mbp\cdot\mbk}{p^{\tau}}-\frac{p_{\|}}{\tau}\partial_{p_{\|}}\right) \delta f_{\mbk,a}
=-\delta C_a[\bar{f},\delta f]\;.
\end{eqnarray}
From this evolution, one can define non-equilibrium Green's functions for moments of the distribution function. In this work, we assume that the background has a vanishing net-charge density, and therefore, the Green function for the net-charge current in position space is
\begin{equation}
\begin{split}
    J^\mu\left( \tau_{\text{hydro}}, \mbx \right) = \int d^2 \mbx^\prime F^{\mu}_{\alpha}\left( |\mbx-\mbx^\prime|,\tau_{\text{hydro}},\tau_{\text{kin}}\right) \delta J^{\alpha}_\mbx\left( \tau_{\text{kin}},\mbx^\prime \right) 
\end{split}
\end{equation}
where $\tau_{\text{kin}}$ is the initial time for the kinetic evolution and $\tau_{\text{hydro}}$ is the time at which hydrodynamic evolution can begin.
The charge-current response functions are calculated by a universal ratio of charge-current vectors from the linear response in kinetic theories. The response functions can then be broken into charge response to charge perturbations (i.e. scalar-scalar), $F_{s}^{s}$, and current response to charge perturbations (i.e. vector-scalar), $F_{s}^{v}$.
\begin{eqnarray}
F_s^s(\tau)=\int \frac{d^2\mbk}{(2\pi)^2} e^{i\mbk\cdot(\mbx-\mbx')}\frac{\tau\delta J_{\rm EKT}^{\tau}(\tau,\mbk)}{\tau_{\rm kin}\delta J^{\tau}(\tau_{\rm kin},\mbk)},~~~~~~~~
F_s^v(\tau)=\int \frac{d^2\mbk}{(2\pi)^2} e^{i\mbk\cdot(\mbx-\mbx')}\frac{i\mbk^i}{|\mbk|}\frac{\tau\delta J_{\rm EKT}^{i}(\tau,\mbk)}{\tau_{\rm kin}\delta J^{\tau}(\tau_{\rm kin},\mbk)}.
\end{eqnarray}


\section{K\o{}MP\o{}ST Framework with QCD Kinetic Theory}
\label{sec:framwork}
\begin{figure*}
\centering
\includegraphics[width=.7\linewidth, height=4cm]{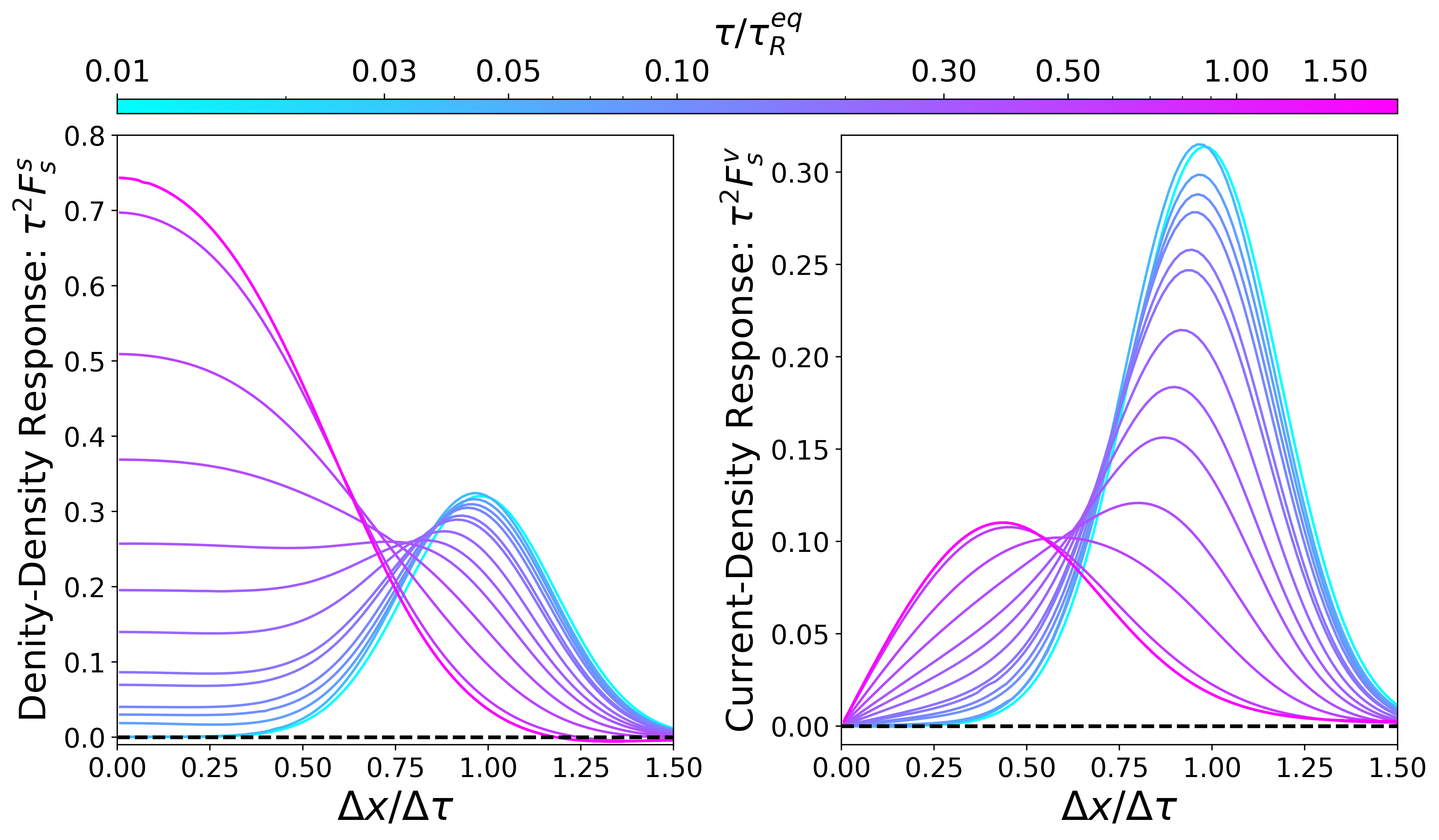}
\caption{Coordinate space Green functions in the charge sector. Eventually, for $\tau/\tau_{R}^{\rm eq} \gtrsim 1$ one sees the emergence of diffusive behavior, as signified e.g. by the hallmark Gaussian structure in the scalar sector.}
\label{fig-chResp}       
\end{figure*}

In Fig.\ \ref{fig-chResp} we show the results of the Green functions for the scalar charge perturbations. In these figures, the black dashed line is a line at $0$ so one can see the Green functions going to $0$ past the causal limit, $\Delta x/\Delta t > 1$, and the finite value of the Green functions past it is a result of smoothing in $k$-space which is required for the Fourier transform. Blue and purple curves in these plots correspond again to early and late times respectively. At early times, the free streaming peak can be clearly seen at $\Delta x/\Delta t =1$. In principle, this should be a delta function-like spike at $\Delta x/\Delta t =1$, but due to the smoothing, becomes a more smoothed-out peak similar to a Gaussian. At late times, the Green functions fall to $0$ at smaller $\Delta x/\Delta t$, which is consistent with the importance of low $k$ response at late times.


We can also offer a physical interpretation of the spatial charge response functions which are shown in Fig.\ \ref{fig-chResp}. For the charge response functions, it is easy to see the emergence of the hydrodynamic limit, especially in the case of the scalar response (the left plot in Fig.\ \ref{fig-chResp}). The late-time behavior displays Gaussian behavior (which is symmetric about $\Delta x$) which is a hallmark of diffusive behavior. On the right side of Fig.\ \ref{fig-chResp}, for the vector response, one can see that the free streaming peak gradually shifts to smaller $\Delta x/\Delta \tau$ and turns into a short distance peak with a longer tail. This is consistent with the late time Navier-Stokes limit, where the vector response for the charge current (i.e. the diffusive current) is given by the gradient of the density (i.e. the gradient of the left side of Fig.\ \ref{fig-chResp}).

\section{Event-by Event Evolution}
\label{sec:EbE}

\begin{figure*}
\centering
\setlength\extrarowheight{10pt}
\begin{tabular}{c}
     \includegraphics[width=.9\linewidth, height=5cm]{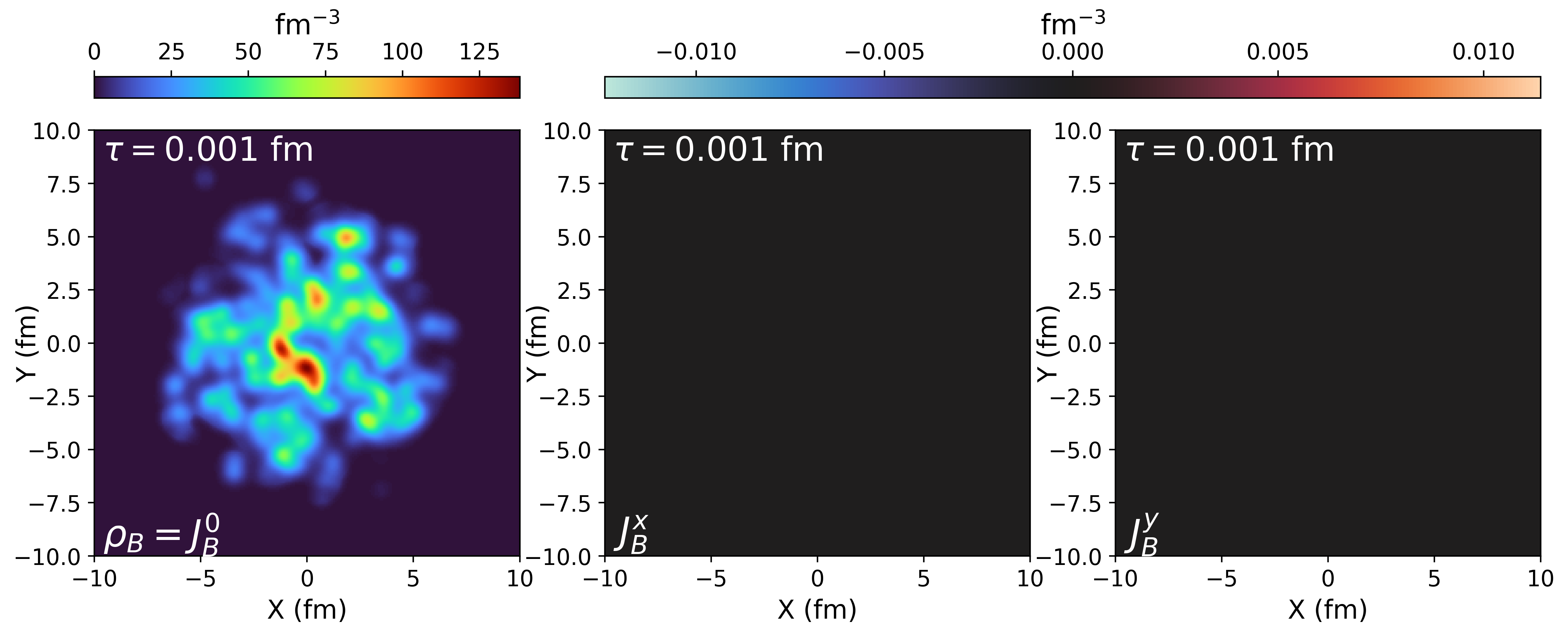} \\
     \includegraphics[width=.9\linewidth, height=5cm]{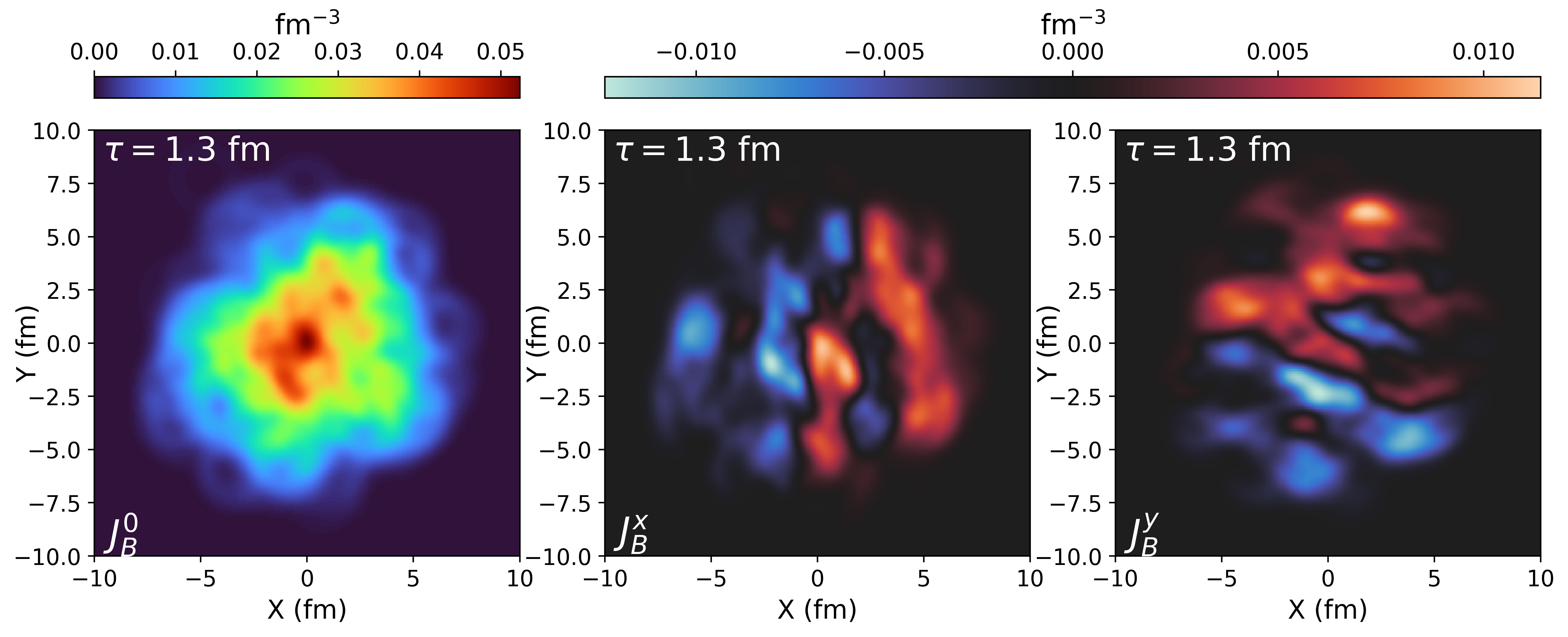} 
\end{tabular}
\caption{The output of the K\o{}MP\o{}ST framework in the charge sector. At late times, the system should be approaching its Navier-Stokes limit.}
\label{fig-jOUT}       
\end{figure*}

In the top row of Fig.\ \ref{fig-jOUT} we show an example of an initial state in the charge sector as provided from the McDIPPER framework~\cite{Garcia-Montero:2023gex,Garcia-Montero:2023opu}. While the McDIPPER framework works in the flavor basis of up, down, and strange, we show results converted to the baryon (B), strange (S), and electric (Q) charge basis. Here, we also only show results for the baryon charge. The hotspots in in Fig.\ \ref{fig-jOUT} track hotspots in the energy density that also exist in the initial condition. The fluctuation scale in this initial state is of the order of the nucleon size, as determined by the initial state model, while the $x$ and $y$ components of the charge vector are initialized to $0$, assuming no charge diffusion at early times.


In the bottom row of Fig.\ \ref{fig-jOUT} we show the results of the K\o{}MP\o{}ST evolution in the charge sector, where the system was evolved to a $\tau_{\text{hydro}}$ of $1.3  {\rm fm}/c$. One can see in the leftmost panel how the gradients in the initial distribution of $J^0$ have been smoothed out. In the rightmost panels, one can see the development of spatial charge current and therefore transverse charge flow. At late times this should approach its hydrodynamic limit, but we leave this detailed analysis for full manuscript \cite{inPrep}




\section{Conclusions and Outlook}
We have upgraded the K\o{}MP\o{}ST framework, which evolves a far-from-equilibrium system towards its hydrodynamic limit, to include conserved charges. By investigating the underlying charge response functions, we find a smooth transition between early free-streaming to hydrodynamic behavior on time scales $\tau/\tau_{R}^{\rm eq} \sim 1$. In the future, we look to include a background for the conserved charges and extend the formalism to $3+1$D.

\section*{Acknowledgements}
This work is supported by the Deutsche
Forschungsgemeinschaft (DFG) under grant CRC-TR 211
“Strong-interaction matter under extreme conditions”
project no. 315477589-TRR 211. XD is also supported
by Xunta de Galicia (Centro singular de investigacion de
Galicia accreditation 2019-2022), European Union ERDF,
the “Maria de Maeztu” Units of Excellence program under
project CEX2020-001035-M, the Spanish Research State
Agency under project PID2020-119632GB-I00, and European Research Council under project ERC-2018-ADG835105 YoctoLHC. The authors gratefully acknowledge the computational resources supported by LUMI-C supercomputer, under The European High-Performance Computing Joint Undertaking grant EHPC-REG-2022R03-192
``Non-equilibrium Quark-Gluon Plasma''; and the National Energy Research Scientific Computing Center (NERSC), a DOE Office of Science User Facility supported by
the Office of Science of the U.S. Department of Energy
under Contract No. DE-AC02-05CH11231.


%
%
%
%
%
\bibliography{ref}

\begin{thebibliography}{12}

\bibitem{Luzum:2008cw}
M.~Luzum, P.~Romatschke, Phys. Rev. C \textbf{78}, 034915 (2008), [Erratum: Phys.Rev.C 79, 039903 (2009)], \texttt{0804.4015}

\bibitem{Giacalone:2017dud}
G.~Giacalone, J.~Noronha-Hostler, M.~Luzum, J.Y. Ollitrault, Phys. Rev. C \textbf{97}, 034904 (2018), \texttt{1711.08499}

\bibitem{ALICE:2015juo}
J.~Adam et~al. (ALICE), Phys. Rev. Lett. \textbf{116}, 222302 (2016), \texttt{1512.06104}

\bibitem{Kurkela:2018wud}
A.~Kurkela, A.~Mazeliauskas, J.F. Paquet, S.~Schlichting, D.~Teaney, Phys. Rev. Lett. \textbf{122}, 122302 (2019), \texttt{1805.01604}

\bibitem{Kurkela:2018vqr}
A.~Kurkela, A.~Mazeliauskas, J.F. Paquet, S.~Schlichting, D.~Teaney, Phys. Rev. C \textbf{99}, 034910 (2019), \texttt{1805.00961}

\bibitem{ALICE:2019nbs}
S.~Acharya et~al. (ALICE), Phys. Lett. B \textbf{807}, 135564 (2020), \texttt{1910.14396}

\bibitem{Garcia-Montero:2023gex}
O.~Garcia-Montero, H.~Elfner, S.~Schlichting (2023), \texttt{2308.11713}

\bibitem{Du:2020dvp}
X.~Du, S.~Schlichting, Phys. Rev. D \textbf{104}, 054011 (2021), \texttt{2012.09079}

\bibitem{Du:2020zqg}
X.~Du, S.~Schlichting, Phys. Rev. Lett. \textbf{127}, 122301 (2021), \texttt{2012.09068}

\bibitem{Arnold:2002zm}
P.B. Arnold, G.D. Moore, L.G. Yaffe, JHEP \textbf{01}, 030 (2003), \texttt{hep-ph/0209353}

\bibitem{Garcia-Montero:2023opu}
O.~Garcia-Montero, S.~Schlichting, H.~Elfner (2023), \texttt{2311.03125}

\bibitem{inPrep}
T.~Dore, X.~Du, S.~Schlichting (In Prep.)

\end{thebibliography}

\end{document}